\documentclass[10pt,conference]{IEEEtran}
\IEEEoverridecommandlockouts
\usepackage{cite}
\usepackage{float}
\usepackage{amsthm,amsmath,amssymb,amsfonts}
\usepackage{mathtools}
\usepackage{algorithmic}
\usepackage{graphicx}
\usepackage{subcaption}
\usepackage{textcomp}
\usepackage{xcolor}
\usepackage{tabularx,ragged2e,booktabs}
\usepackage{multirow}
\usepackage{rotating}
\usepackage[utf8]{inputenc}
\usepackage{dblfloatfix}
\usepackage{ulem}
\usepackage[hang,flushmargin]{footmisc}
\usepackage[font=small,labelfont=bf]{caption}
\usepackage{url}

\newtheorem{definition}{Definition}
\newtheorem{theorem}{Theorem}
 
\def\BibTeX{{\rm B\kern-.05em{\sc i\kern-.025em b}\kern-.08em
    T\kern-.1667em\lower.7ex\hbox{E}\kern-.125emX}}

\begin{document}

\title{Empirical Analysis of Randomness Quality in Differential Privacy Mechanisms}

\author{
    \IEEEauthorblockN{Cesare Gerolimetto Fabrello\IEEEauthorrefmark{1}\IEEEauthorrefmark{2}, Valeria Rossi\IEEEauthorrefmark{1}\IEEEauthorrefmark{2}, Alberto Trombetta\IEEEauthorrefmark{1}, and Massimo Caccia\IEEEauthorrefmark{1}\IEEEauthorrefmark{2}}
    \IEEEauthorblockA{\IEEEauthorrefmark{1}Università degli Studi dell'Insubria}
    \IEEEauthorblockA{\IEEEauthorrefmark{2}Random Power Srl}
}

\maketitle

\begin{abstract}
Differential Privacy (DP) relies on carefully calibrated random noise to protect individual privacy in statistical analyses. While theoretical work has analyzed DP under weakened randomness assumptions, the practical consequences of entropy degradation remain poorly understood. We present a systematic empirical investigation of how randomness quality affects differential privacy mechanisms using IBM's DiffPrivLib. We introduce progressively degraded entropy sources characterized by established test suites, starting from high-quality quantum True Random Number Generators (TRNGs) and cryptographically secure Pseudo-Random Number Generators (PRNGs) down to systematically manipulated sources with controlled entropy degradation. Through repeated experiments over one million queries on a reference database and complementary statistical tests, we directly analyze empirical Privacy Loss Random Variable distributions. Our results demonstrate that DP mechanisms reliably detect deviations when approximately 1 bit in every 8 to 16 is manipulated, with detection sensitivity varying significantly between bit-level biases and temporal correlations. We demonstrate that statistical detection of distributional anomalies does not necessarily correspond to actual privacy guarantee violations.
\end{abstract}


\thispagestyle{plain}
\pagestyle{plain}
\pagenumbering{gobble}

\section{Introduction}
\label{sec:intro}
Randomness is fundamental to modern security and information systems, enabling encryption key generation, secure data transmission, and information protection \cite{gennaro}. The quality of randomness directly determines whether cryptographic protocols function as intended. Notable failures demonstrate severe consequences of weak entropy: the Debian OpenSSL vulnerability (2008) compromised millions of cryptographic keys \cite{debian2}, while the \texttt{Dual\_EC\_DRBG} backdoor revealed how weakened randomness could undermine entire security infrastructures \cite{shumow2007dualec, nytimes2013backdoors, nist2014dualec}.

Traditional anonymization methods (removing direct identifiers) have proven insufficient, as Narayanan and Shmatikov demonstrated that even few anonymous data points enable re-identification when combined with publicly available information \cite{narayanan2007breakanonymitynetflixprize}. Differential Privacy (DP) has emerged as a mathematically rigorous framework addressing this challenge \cite{dp_book}. It ensures that analysis results barely change whether any specific individual's data is included, by adding carefully calibrated random noise to computation results, with the amount governed by a privacy loss parameter.

However, implementation on finite computers introduces vulnerabilities. Notably, Mironov demonstrated that floating-point representation can leak information \cite{mironov_lsb}.

The quality of randomness used in DP mechanisms has sparked particular debate. The U.S. Census Bureau's adoption of Differential Privacy for protecting 2020 census data marked one of the first large-scale governmental applications of the framework \cite{us-census}. In analyzing their Disclosure Avoidance System, concerns were raised that using pseudo-random number generators might be insufficient, with concerns that ``the information-theoretic privacy-loss budget may be larger than claimed" when using pseudo-random number generators \cite{us-census}. Conversely, theoretical work has shown that DP can be preserved even under imperfect randomness sources, such as Santha-Vazirani sources where each bit retains only partial unpredictability, provided non-additive mechanisms are employed \cite{imperfect_dp}.

\underline{\textbf{Our contributions:}} While prior work has analyzed differential privacy under weakened randomness assumptions from a theoretical standpoint, the practical consequences of entropy degradation in concrete DP implementations remain empirically uncharacterized. We provide a systematic empirical investigation of how randomness quality influences differential privacy mechanisms when executed on finite computers. We design a controlled experimental framework based on IBM's DiffPrivLib \cite{ibm_diffprivlib}, evaluating canonical DP mechanisms under fixed privacy budgets, query types, and numerical parameters while systematically varying the entropy source driving noise generation. We introduce progressively degraded random sources whose statistical properties are independently characterized using established test suites, starting from high-quality quantum TRNGs and cryptographically secure PRNGs down to systematically manipulated sources with controlled entropy degradation.

Rather than inferring privacy guarantees indirectly, we directly analyze empirical output distributions on neighboring datasets through repeated executions with large sample sizes. We apply complementary statistical tests (including sign tests, sigma exceedance analyses, and chi-squared goodness-of-fit tests) to detect deviations from the theoretical model and quantify whether entropy degradation produces statistically detectable biases in the Privacy Loss Random Variable, even when mechanisms remain theoretically differentially private.

The paper is organized as follows: Section \ref{sec:dp} introduces the mathematical framework of Differential Privacy and its main mechanisms. Section \ref{sec:materials_and_methods} describes our experimental methodology, including the characterization of entropy sources and statistical testing framework. Section \ref{sec:model_validation} presents model validation results. Section \ref{sec:results} analyzes the impact of entropy quality on differential privacy mechanisms through complementary statistical tests. Section \ref{sec:conclusions} concludes with implications for practical implementations.

\section{Differential Privacy: Mathematical Framework}
\label{sec:dp}

Differential Privacy establishes a mathematically rigorous framework that overcomes the drawbacks of traditional approaches. We begin by defining what constitutes a Randomized Algorithm:

\begin{definition}[Randomized Algorithm\cite{dp_book}]
A Randomized Algorithm $\mathcal{M}$ with domain $A$ and range $B$ is an algorithm associated with a total map $M:A\rightarrow \Delta (B)$. On input $a\in B$, the algorithm $\mathcal{M}$ outputs $\mathcal{M}(a) = b$ with probability $(M(a))_b$ for each $b \in B$.
\end{definition}

The algorithm is non-deterministic in nature, as computations require harvesting randomness to proceed, and the output probability space is over the random choices made by the mechanism.

\begin{definition}[Differential Privacy\cite{dp_book}]
A randomized algorithm $\mathcal{M}$ is ($\varepsilon$, $\delta$)-differentially private if, for all neighboring datasets $\mathcal{D}$ and $\mathcal{D}'$ (i.e. that differ by at most one element), and for all measurable subsets $\mathcal{S}$ of the output space of $\mathcal{M}$:

\begin{equation}
P(\mathcal{M}(\mathcal{D}) \in \mathcal{S}) \leq e^{\varepsilon} \cdot P(\mathcal{M}(\mathcal{D}') \in \mathcal{S}) + \delta
\end{equation}

where $\varepsilon$ is the privacy loss (or privacy budget) parameter and $\delta$ accounts for a small probability of failing the privacy guarantees.
\end{definition}

Informally, this definition ensures that given two neighboring datasets fed as input to the mechanism, the distributions of outputs are close, where this closeness is parameterized by the privacy loss $\varepsilon$. A smaller $\varepsilon$ provides stronger privacy guarantees by requiring the output distributions to be nearly identical, at the cost of reduced accuracy in query results due to increased noise. The strongest form occurs when $\delta = 0$, referred to as Pure Differential Privacy, providing tight bounds on $\varepsilon$. When $\delta > 0$, we have Approximate Differential Privacy, which accounts for a small probability of privacy guarantee failure but requires less noise for the same privacy level.

To measure the distance between output distributions on neighboring datasets, we introduce the Privacy Loss Random Variable (PLRV). For a mechanism $\mathcal{M}$ and neighboring datasets $\mathcal{D}, \mathcal{D}'$, the privacy loss at output $y$ is:

\[
L_{\mathcal{M}}^{\mathcal{D},\mathcal{D}'}(y) = \ln \left( \frac{P(\mathcal{M}(\mathcal{D}) = y)}{P(\mathcal{M}(\mathcal{D}') = y)} \right).
\]

A mechanism is $\varepsilon$-DP exactly when the privacy loss is almost surely bounded by $\varepsilon$, and $(\varepsilon,\delta)$-DP when this bound holds with probability at least $1-\delta$. The PLRV therefore provides a precise tool to analyze privacy guarantees.

\subsection{Differential Privacy Mechanisms}

The most commonly used mechanisms for implementing differentially private numerical queries are the Laplace and Gaussian mechanisms.

\begin{definition}[Laplace Mechanism\cite{dp_book}]
Given a function $f : \mathcal{D} \rightarrow \mathbb{R}$, the Laplace mechanism for $f$ with scale $\lambda$ is defined as
\begin{equation}
\mathcal{M}(u) = f(u) + Y; \ \text{where} \ Y \sim Lap(\lambda)
\end{equation}
where $Lap(\lambda)$ denotes the Laplace probability distribution centered at zero, defined by the probability density function 
\begin{equation}
g(\lambda ; x) = \frac{1}{2\lambda}e^{-\frac{|x|}{\lambda}}
\end{equation}
\end{definition}

For a query function $f$, the mechanism returns $f(\mathcal{D}) + Lap(\Delta f / \varepsilon)$, where $\Delta f$ represents the sensitivity of function $f$:

\begin{equation}
\Delta f = \sup_{D, D'} | f(D) - f(D') |
\end{equation}

for all neighboring datasets $\mathcal{D}$ and $\mathcal{D}'$. This sensitivity analysis determines the required noise amount: queries with higher sensitivity need proportionally more noise for the same privacy protection. It can be proven that: 

\begin{theorem}
The Laplace mechanism preserves $\varepsilon$-DP.
\end{theorem}

The Gaussian mechanism similarly adds noise sampled from a normal distribution $\mathcal{N}(0, \sigma^2)$, where the noise scale $\sigma$ must satisfy specific bounds relative to the L2-sensitivity of the query. While it achieves only approximate $(\varepsilon, \delta)$-DP due to the non-zero failure probability $\delta$, it often requires less noise than Laplace for the same privacy level, improving statistical utility \cite{dp_book}.

\subsection{Discrete Mechanisms}

Real-world implementations often use discrete versions of these mechanisms, as query outputs are frequently integer-valued and discrete sampling avoids floating-point numerical errors.

\begin{definition}[Discrete Laplace Distribution \cite{discrete_laplace1}]
Given parameter $p = e^{-1/\sigma}$, where $\sigma >0$ is the scale, a random variable $\mathcal{Y}$ has the discrete Laplace distribution DL(p) if 
\begin{equation}
f_p(k) = P(\mathcal{Y} = k) = \frac{1-p}{1+p}\, p^{|k|}, \qquad k \in \mathbb{Z}
\end{equation}
\end{definition}

\begin{definition}[Discrete Laplace Mechanism]
Given a function $f : \mathcal{D} \rightarrow \mathbb{Z}$, privacy parameter $\varepsilon$, and sensitivity $\Delta s$, the discrete Laplace mechanism outputs
\begin{equation}
\mathcal{M}(u) = f(u) + Y, \quad \text{where} \ Y \sim \text{DL}(\Delta s / \varepsilon)
\end{equation}
\end{definition}

The discrete Laplace mechanism satisfies pure $\varepsilon$-DP and guarantees integer-valued noise, avoiding floating-point precision loss. The Discrete Gaussian mechanism provides a similar integer-valued alternative for approximate-DP applications, discretizing the continuous Gaussian distribution onto the integers while maintaining comparable privacy guarantees.

While discrete mechanisms address concerns about integer outputs and numerical stability, they do not fully resolve implementation challenges. Even mathematically correct mechanisms face vulnerabilities from the gap between theory and practice, which can compromise privacy guarantees.

\begin{figure*}[htp]
	\centering
	\begin{subfigure}[b]{0.45\textwidth}
		\centering
		\includegraphics[width=\textwidth]{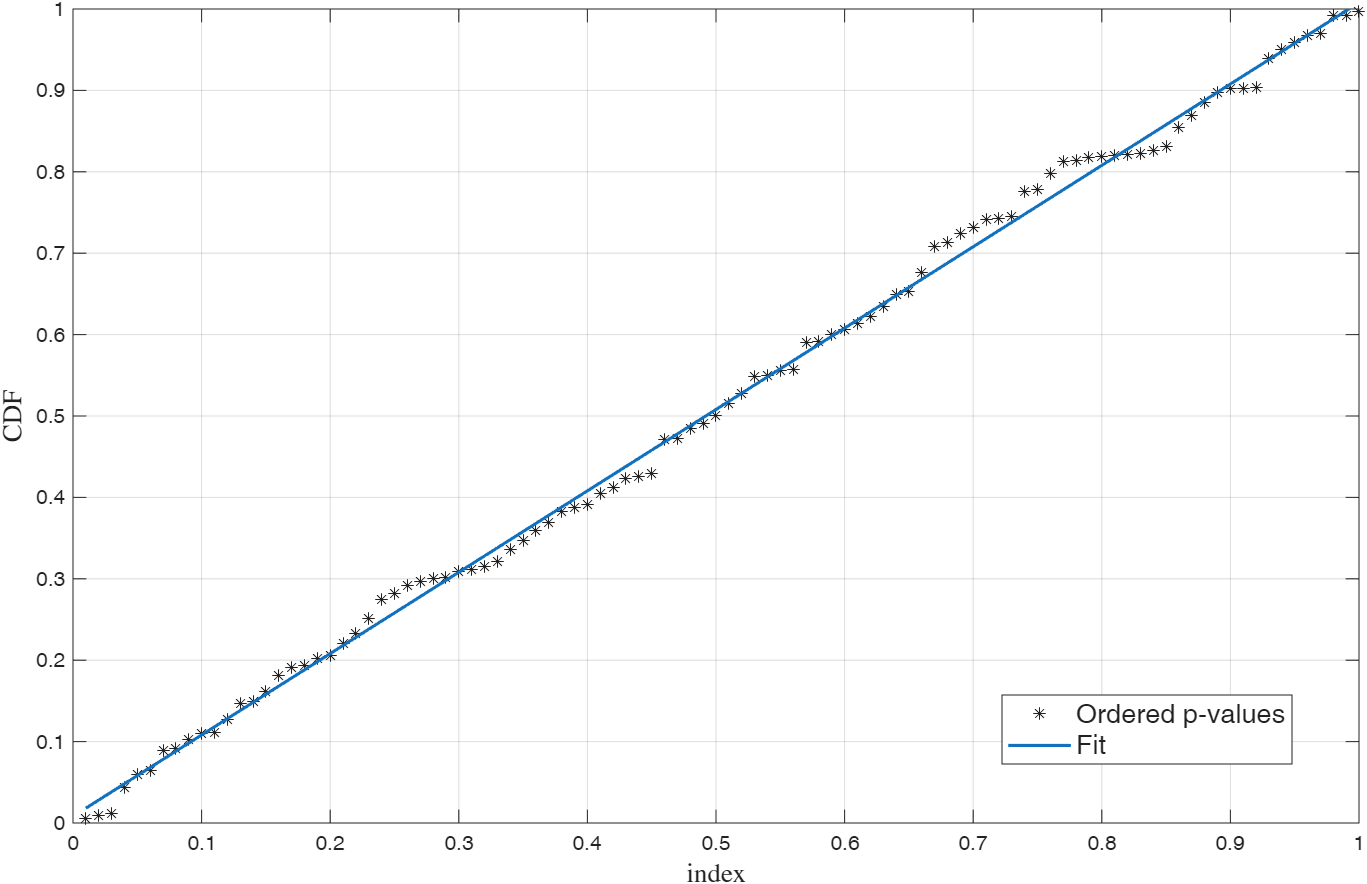}
		\caption{Cumulative distribution of p-values.}
		\label{fig:laplace_pvals_cdf}
	\end{subfigure}
	\hfill
	\begin{subfigure}[b]{0.45\textwidth}
		\centering
		\includegraphics[width=\textwidth]{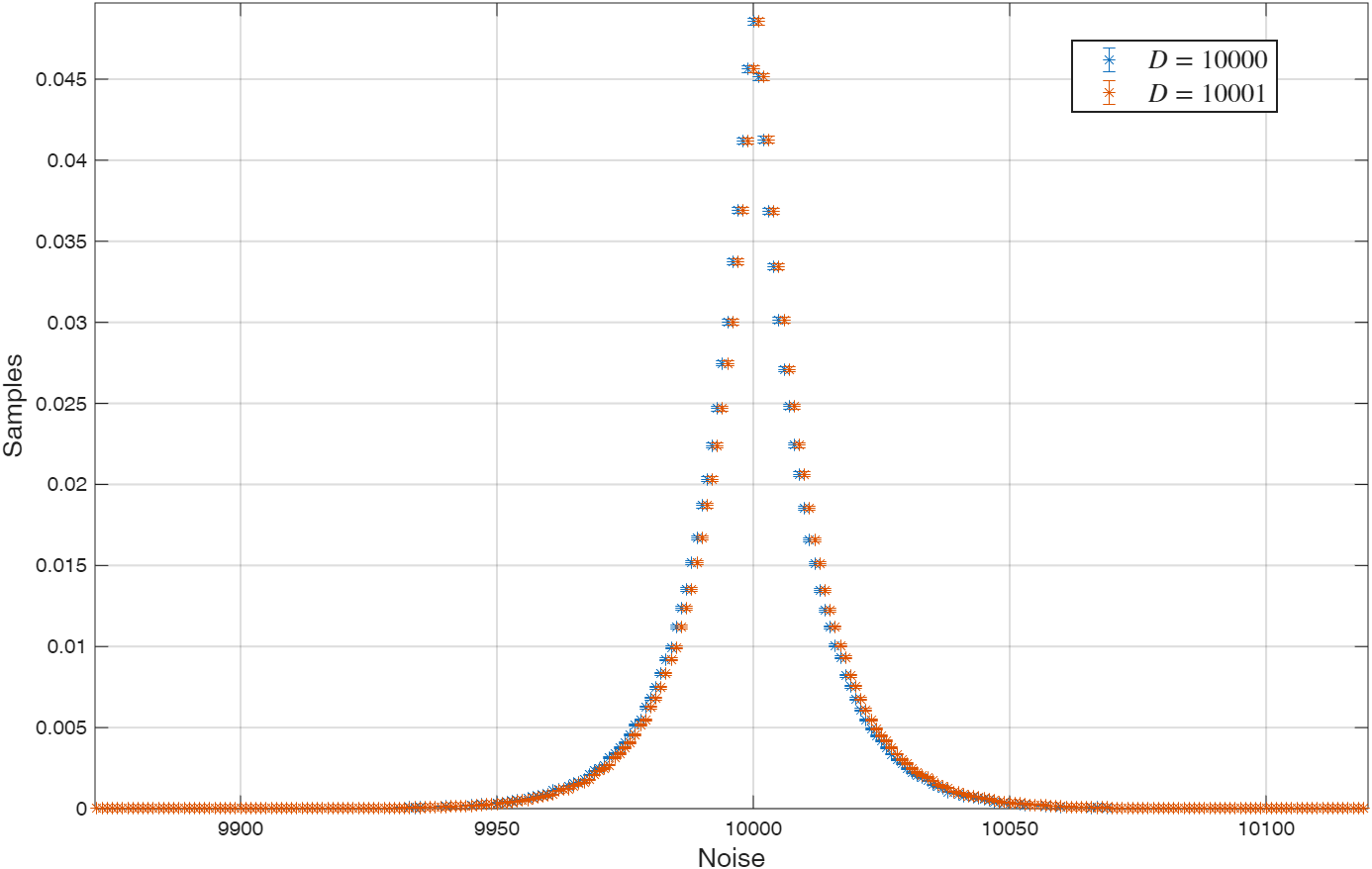}
		\caption{Output histograms from neighboring datasets.}
		\label{fig:ibm_rap_hists}
	\end{subfigure}
	\begin{subfigure}[b]{0.45\textwidth}
		\centering
		\includegraphics[width=\textwidth]{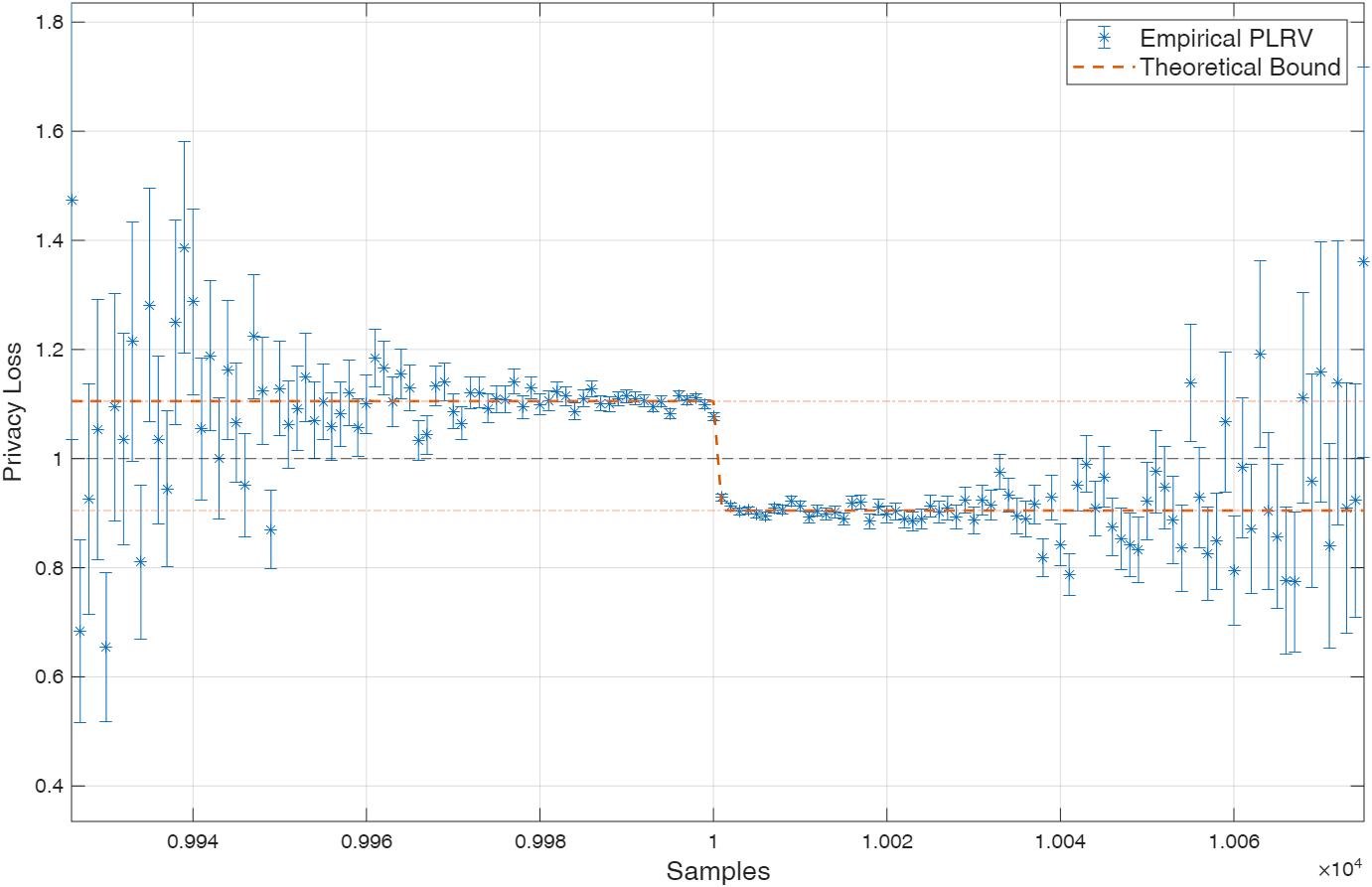}
		\caption{Empirical PLRV distribution ($n_{bins} = 150$).}
		\label{fig:ibm_rap_plrv}
	\end{subfigure}
	\hfill
	\begin{subfigure}[b]{0.45\textwidth}
		\centering
		\includegraphics[width=\textwidth]{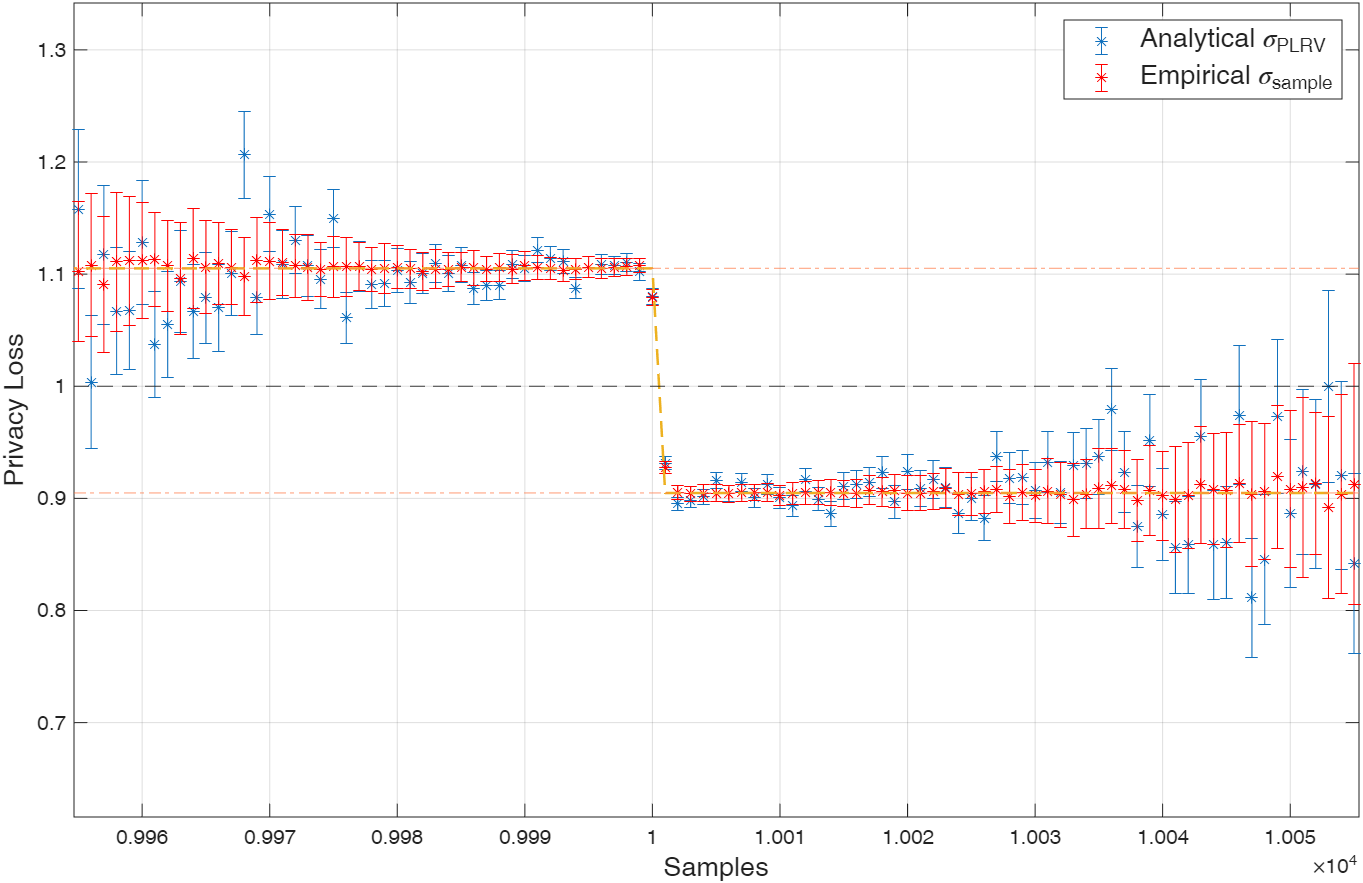}
		\caption{Uncertainty model validation.}
		\label{fig:statistical_significance}
	\end{subfigure}
    \caption{Model validation and uncertainty quantification. (a) Cumulative distribution of p-values from 100 independent $\chi^2$ tests shows the expected linear trend (fitted parameters $a = 0.999 \pm 0.005$ and $b = 0.008 \pm 0.003$). (b) Output histograms from queries on $\mathcal{D}_1$ (blue, centered at 10,000) and $\mathcal{D}_2$ (orange, centered at 10,001) using \texttt{REF} entropy source. (c) Corresponding PLRV computed from the histograms in (b). (d) Ratio of analytical uncertainty ($\sigma_{\text{PLRV}}$, blue) to empirical standard deviation ($\sigma_{100}$, orange) across bins. Values near 1 confirm the validity of the binomial error model. Larger deviations appear in the tails where bin populations are small and statistical noise dominates.}
	\label{fig:cdf_plrv_statistical_significance}

    \vspace{-1.5em}
\end{figure*}

\section{Materials and Methods}
\label{sec:materials_and_methods}

The aim of this analysis is to compute the empirical PLRV by generating samples through repeated experiments and applying the definition of differential privacy directly. We compare the empirical PLRV with the theoretical distribution predicted by the $\varepsilon$-DP model, measuring statistically significant deviations that arise from varying randomness sources with gradually degrading entropy quality. All experiments use the Discrete Laplace mechanism.

All random data were pre-generated, stored in binary files, and fed directly to IBM's DiffPrivLib by modifying its native random number generator. All experiments ran using IBM's DiffPrivLib v.0.6.6 on dual Intel Xeon Gold 5416S processors with 128 GB RAM; post-processing was performed in MATLAB. Codebase is available on a Github repository\footnote{\url{https://github.com/grlcsr/dp_analysis}}.

\subsection{Randomness Sources}

We employ multiple randomness sources with varying statistical properties, independently qualified using well-known test suites. The bit-streams are saved into files of 2 Gbit each, gathered in sets of 20 files for a total of 40 Gbit.

\subsubsection{Entropy Sources}

\begin{itemize}
    \item \textbf{REF (Reference) \cite{rap!}}: A well-calibrated QRNG (developed by Random Power\footnote{https://www.randompower.eu/}) serving as our baseline. This source was proven to provide streams of unpredictable, independent and identically distributed symbols.
    
    \item \textbf{CSPRNG \cite{openssl_rand}}: OpenSSL's cryptographically secure pseudo-random number generator. Despite being deterministic, CSPRNGs are designed to be computationally indistinguishable from true randomness and should pass statistical tests in a similar manner to \texttt{REF}. 
    
    \item \textbf{MIS (Miscalibrated)}: A deliberately mis-calibrated QRNG producing subtle correlations in the bit-stream. 
    
    \item \textbf{Controlled manipulations (M4, M8, M16, M32, M64)}: To systematically study entropy degradation, we introduce reproducible biases by partitioning the \texttt{REF} bit-stream into consecutive symbols of size $b \in \{4, 8, 16, 32, 64\}$ bits and applying exactly one deterministic bit operation per symbol. The manipulation frequency decreases from M4 (most frequent, every nibble) to M64 (most sparse, every 8 bytes).
\end{itemize}

\subsubsection{Manipulation Types}

We employ three manipulation processes that introduce distinct forms of non-randomness:

\begin{enumerate}
    \item \textbf{FLIP}: For each symbol of $b$ bits length and integer value $x$, compute target bit index $p = x \bmod b$ and flip bit $p$. This creates moderate per-symbol distortions in the bit distribution.
    
    \item \textbf{ALTERNATE}: Similar to FLIP but alternates between clearing and setting bit $p$ on successive symbols (with $p$ still determined by current symbol). This introduces both spatial and temporal patterns.
    
    \item \textbf{CORRELATION}: For each symbol, compute $p = x_{\text{prev}} \bmod b$ using the previous symbol's value, then copy the bit value from position $p$ of the previous symbol into position $p$ of the current symbol. This introduces explicit temporal dependencies between consecutive symbols.
\end{enumerate}

Two additional manipulation types, defined as \textbf{CLEAR} (deterministically clearing bit $p$) and \textbf{SET} (deterministically setting bit $p$), were also tested but proved too catastrophic for meaningful analysis. By deterministically clearing or setting one bit in each symbol, these manipulations drastically skew the symbol distribution: some symbol values become artificially over-represented while others are suppressed or disappear entirely. These manipulations consistently failed test batteries even at the lowest manipulation frequency (M64), making them unsuitable for establishing sensitivity gradients. 

\subsubsection{Test Batteries and Qualification Framework}
We validated entropy sources using NIST SP 800-22 \cite{nist800-22}, TestU01 \cite{testu01}, and NIST SP 800-90B IID validation \cite{nist800-90b}, which assess temporal structure, correlations, and symbol-level independence respectively. Entropy estimation is performed on 4-bit symbols (nibbles) as the reference QRNG produces 4 bits per cycle, making this the natural base unit for our analysis.

\subsubsection{Qualification Results}
Baseline sources (\texttt{REF}, \texttt{CSPRNG}) achieve near-perfect min-entropy ($H_{\min} \geq 3.9995$ bits per 4-bit symbol) and pass all test batteries. \texttt{MIS} maintained high min-entropy but exhibited temporal correlation failures. Controlled manipulations showed progressive entropy degradation (Table \ref{tab:iid_entropy}), with \texttt{CORRELATION} consistently failing IID validation. The gradient from M4 to M64 provides a controlled spectrum of entropy quality, allowing us to identify the sensitivity threshold at which DP mechanisms begin to show statistically detectable deviations from theoretical behavior.

\subsection{Experimental Parameters}
We designed our framework based on counting the number of entries in a dataset (length query), chosen for its simplicity and sensitivity equal to 1. Our analysis focuses on $\varepsilon = 0.1$ (results for other values such as $\varepsilon = 0.01$ showed similar patterns). For neighboring datasets, we used fixed sizes of 10,000 ($\mathcal{D}_1$) and 10,001 ($\mathcal{D}_2$) entries. 

We ran 1 million iterations for each combination of parameters and randomness source. To assess robustness, we varied histogram bin sizes (1, 3, 5) and configurations, and results were consistent. We primarily report bin size 1 with a fixed binning of 75 bins of size 1 per side of the midpoint, ensuring coherent comparisons across all entropy sources and manipulation levels (Figure \ref{fig:cdf_plrv_statistical_significance}).

The datasets used (Adult and US 1990 Census) are publicly available and represent common use cases, though their specific content is irrelevant since our query simply counts individuals.

\subsection{Statistical Metrics}
\label{subsec:metrics}

Measuring the influence of biased randomness on differential privacy requires careful consideration of what we mean by ``influence". Our analysis focuses on two fundamental questions: first, how much bias is needed to produce measurable deviations from the theoretical model in terms of the PLRV? Second, even if the measured output distribution deviates from the expected Laplacian model, does this necessarily mean the privacy guarantee is broken?

Since Laplace sampling procedures expect uniformly distributed random input, we need statistical tools that can detect when this assumption is violated and quantify the resulting impact on output distributions. After evaluating various alternatives, we settled on three standard complementary tests to assess whether the empirical distribution deviates significantly from the theoretical model:

\begin{itemize}
    \item \textbf{Sign test}: A non-parametric test that detects systematic directional biases. The sign test identifies whether the output consistently skews in one direction relative to the theoretical model.
    
    \item \textbf{$k\sigma$ residual exceedance test}: We standardize the residuals $z_i = (O_i - E_i)/\sigma_i$ between the empirical PLRV (observed, $O_i$) and theoretical distribution (expected, $E_i$) for each bin $i$. Under the hypothesis that our samples follow the theoretical model, the residuals should be approximately Gaussian. We then compute the fraction of bins where $|z_i| > k$ and compare this to the expected rate from a normal distribution for varying $k \in \{1, 2\}$. We limited analysis to $k \leq 2$ due to sample size constraints; at $N=10^6$, $3\sigma$ exceedances are too rare (expected rate $\sim$0.3\%)  for reliable statistics with the typical $\sim200$ bins in our histograms. Significant exceedance indicates poor fit of the distribution. These PLRV values are where the privacy guarantee might fail more likely, which means a higher risk of outputs that reveal more than the stated privacy budget should permit.
    
    \item \textbf{$\chi^2$ goodness-of-fit test}: Evaluates whether the observed PLRV distribution matches the expected theoretical distribution. This test is sensitive to overall shape differences and provides a global measure of fit quality across all bins of the distribution.
\end{itemize}

Together, these metrics allow us to quantify how well the empirical PLRV fits the theoretical model and determine the minimum amount of bias needed to produce statistically significant deviations. Importantly, detecting such deviations does not automatically imply privacy failure; rather, it indicates where the gap between theory and practice becomes measurable, allowing us to assess the robustness margins of differential privacy mechanisms under realistic entropy conditions. This standardized setup allows us to make unbiased comparisons that isolate the effect of randomness quality from other factors that could affect the results. By maintaining consistency in data inputs, query formulations, and parameter settings, along with computational environment (which eliminates hardware-related variations), we can focus on observing potential differences in output distributions due to the selected randomness sources.

\begin{table}[t]
\centering
\small
\begin{tabularx}{\columnwidth}{@{}l *{3}{>{\centering\arraybackslash}X} @{}}
\toprule
\textbf{Source} & \textbf{ALTERNATE} & \textbf{FLIP} & \textbf{CORREL.} \\
\midrule
M4      & 3.4148 & 2.9998 & 3.1926 \\
M8      & 3.6779 & 3.4149 & 3.7518 \\
M16     & 3.8301 & 3.6780 & 3.9124 \\
M32     & 3.9503 & 3.9021 & 3.9719 \\
M64     & 3.9861 & 3.9722 & 3.9913 \\
\bottomrule
\end{tabularx}
\caption{Min-entropy estimates (bits per 4-bit symbol) for controlled manipulations, which show progressive entropy recovery from M4 to M64.}
\label{tab:iid_entropy}
\end{table}

\section{Model Validation}
\label{sec:model_validation}

Our first step was to reconstruct the empirical Laplace distributions from collected samples by building histograms. With sensitivity $\Delta s = 1$ and $\varepsilon = 0.1$, the theoretical Laplace scale becomes $\Delta s / \varepsilon = 10$. Histogram reconstruction from $N = 10^6$ samples using bin size 1 accurately reproduced the expected discrete Laplace distribution. To quantify conformity to the theoretical model, we performed $\chi^2$ goodness-of-fit tests, repeating each test 100 times independently to ensure robustness. IBM's DiffPrivLib consistently produced reduced $\chi^2$ values near 1, with p-values showing the expected linear cumulative distribution trend (Figure \ref{fig:laplace_pvals_cdf}), further validated by a Kolmogorov-Smirnov test against the uniform distribution ($p = 0.93$).

\subsection{Privacy Loss Random Variable Computation}

Following the definition of differential privacy directly, we constructed neighboring datasets $\mathcal{D}_1$ and $\mathcal{D}_2$ with 10,000 and 10,001 individuals respectively. For each dataset, we executed $N$ queries and built corresponding output histograms (Figure \ref{fig:ibm_rap_hists}). We computed the empirical PLRV as the ratio of bin populations: $\text{PLRV}_i = n_{1,i}/n_{2,i}$, where $n_{1,i}$ and $n_{2,i}$ are the number of samples in bin $i$ from queries on $\mathcal{D}_1$ and $\mathcal{D}_2$ respectively. Figure \ref{fig:ibm_rap_plrv} shows the resulting PLRV distribution, with values bounded within $e^{\pm\varepsilon}$ in the central region as expected for pure differential privacy with $\varepsilon = 0.1$. 

\subsection{Uncertainty Quantification}

Before analyzing entropy effects, we validated our measurement procedure by establishing that observed variability matches expected sampling statistics. Assuming binomial statistics for histogram populations with $N = 10^6$ samples, we derived analytical uncertainties through error propagation: $\sigma_{\text{PLRV}}^2 = \frac{1}{n_2^2} \sigma_{D_1}^2 + \frac{n_1^2}{n_2^4} \sigma_{D_2}^2$, where $\sigma_{D_1}^2 = n_1(1 - n_1/N)$ and $\sigma_{D_2}^2 = n_2(1 - n_2/N)$.

To validate this analytical model, we repeated each experiment $M = 100$ times and computed the empirical standard deviation $\sigma_{100}$ across repetitions. We then compared the ratio $R = \sigma_{\text{PLRV}}/\sigma_{100}$ for each bin. If our binomial model is correct, we expect $R \approx 1$ for most bins, with deviations confined to distribution tails where bin populations are small.

\begin{table}[t]
\centering
\caption{Sign test results on full PLRV histogram (bin size 1, $n=150$) for baseline entropy sources. $n$ is the number of bins, $k_{\text{pos}}$ is the number of bins with positive deviations, $p$ is the $p$-value, $p^*$ is the $p$-value after excluding the two bins at the PLRV discontinuity, and $H_{\min}$ is the min-entropy.}
\label{tab:sign_test_baseline}
\small
\begin{tabular*}{\columnwidth}{@{\extracolsep{\fill}}lccccc}
\toprule
\textbf{Source} & $n$ & $k_{\text{pos}}$ & $p$ & $p^*$ & $H_{\min}$ \\
\midrule
\texttt{REF}     & 150 & 74 & 0.935 & 1.000 & 3.9995 \\
\texttt{CSPRNG}  & 150 & 69 & 0.369 & 0.460 & 3.9995 \\
\texttt{MIS}     & 150 & 66 & 0.165 & 0.217 & 3.9954 \\ 
\bottomrule
\end{tabular*}

\vspace{0.5em}

\caption{$k\sigma$ exceedance test results on full PLRV histogram (bin size 1, $n=150$) for baseline entropy sources. For each test level, we report the observed exceedance rate, its $p$-value, and $p^*$ (after excluding the two bins at the PLRV discontinuity).}
\label{tab:ksigma_baseline}
\small
\begin{tabular*}{\columnwidth}{@{\extracolsep{\fill}}lcccccc}
\toprule
& \multicolumn{3}{c}{\textbf{1$\sigma$ exceedance}} & \multicolumn{3}{c}{\textbf{2$\sigma$ exceedance}} \\
\cmidrule(lr){2-4} \cmidrule(lr){5-7}
\textbf{Source} & Rate & $p$ & $p^*$ & Rate & $p$ & $p^*$ \\
\midrule
\texttt{REF}     & 0.327 & 0.366 & 0.458 & 0.073 & 0.042 & 0.139 \\
\texttt{CSPRNG}  & 0.347 & 0.194 & 0.264 & 0.073 & 0.042 & 0.139 \\
\texttt{MIS}     & 0.327 & 0.366 & 0.458 & 0.060 & 0.147 & 0.361 \\

\bottomrule
\end{tabular*}

\vspace{0.5em}

\caption{$\chi^2$ goodness-of-fit test results on full PLRV histogram (bin size 1, $n=150$) for baseline entropy sources. $\chi^{2*}_{\text{red}}$ and $p^*$ denote values obtained after excluding the two bins at the PLRV discontinuity. Under the null hypothesis, $\chi^2_{\text{red}} \approx 1$.}
\label{tab:chi2_baseline}
\small
\begin{tabular*}{\columnwidth}{@{\extracolsep{\fill}}lcccc}
\toprule
\textbf{Source} & $\chi^2_{\text{red}}$ & $p$ & $\chi^{2*}_{\text{red}}$ & $p^*$ \\
\midrule
\texttt{REF}     & 1.30 & 0.008 & 1.11 & 0.174 \\
\texttt{CSPRNG}  & 1.31 & 0.007 & 1.09 & 0.207 \\
\texttt{MIS}     & 1.15 & 0.102 & 0.95 & 0.667 \\
\bottomrule
\end{tabular*}
\end{table}

\begin{table*}[htbp]
\centering
\caption{Sign test results on full PLRV histogram (bin size 1) for controlled entropy manipulations. For each manipulation level and type, we report $n$ (number of bins), $k_{\text{pos}}$ (number of bins with positive deviations), $p$ (two-sided $p$-value). $p^*$ denotes the $p$-value obtained after excluding the two bins at the PLRV discontinuity.}
\label{tab:sign_test_manipulation}
\small
\begin{tabular*}{\textwidth}{@{\extracolsep{\fill}}l*{4}{c}*{4}{c}*{4}{c}}
\toprule
& \multicolumn{4}{c}{\textbf{FLIP}} & \multicolumn{4}{c}{\textbf{ALTERNATE}} & \multicolumn{4}{c}{\textbf{CORRELATION}} \\
\cmidrule(lr){2-5} \cmidrule(lr){6-9} \cmidrule(lr){10-13}
\textbf{Level} 
& $n$ & $k_{\text{pos}}$ & $p$ & $p^*$ 
& $n$ & $k_{\text{pos}}$ & $p$ & $p^*$ 
& $n$ & $k_{\text{pos}}$ & $p$ & $p^*$ \\
\midrule
\texttt{M4}  & 150 & 36  & 0     & 0     & 147 & 64  & 0.137 & 0.135 & 125 & 100 & 0     & 0     \\
\texttt{M8}  & 150 & 57  & 0.004 & 0.004 & 150 & 66  & 0.165 & 0.217 & 150 & 85  & 0.121 & 0.118 \\
\texttt{M16} & 150 & 61  & 0.027 & 0.040 & 150 & 71  & 0.568 & 0.681 & 150 & 74  & 0.935 & 1.000 \\
\texttt{M32} & 150 & 68  & 0.289 & 0.366 & 150 & 74  & 0.935 & 1.000 & 150 & 70  & 0.463 & 0.565 \\
\texttt{M64} & 150 & 80  & 0.463 & 0.366 & 150 & 79  & 0.568 & 0.460 & 150 & 71  & 0.568 & 0.681 \\
\bottomrule
\end{tabular*}

\vspace{0.5em}

\caption{$k\sigma$ exceedance test results on full PLRV histogram (bin size 1, $n=150$) for controlled entropy manipulations. For each manipulation level, we report the observed exceedance rate and its $p$-value. $p^*$ denotes the $p$-value obtained after excluding the two bins at the PLRV discontinuity. Expected rates under the null hypothesis are 31.7\% ($1\sigma$) and 4.6\% ($2\sigma$).}
\label{tab:ksigma_manipulated}
\scriptsize
\begin{tabular*}{\textwidth}{@{\extracolsep{\fill}}l*{3}{ccc}*{3}{ccc}}
\toprule
& \multicolumn{9}{c}{\textbf{1$\sigma$ exceedance}} & \multicolumn{9}{c}{\textbf{2$\sigma$ exceedance}} \\
\cmidrule(lr){2-10} \cmidrule(lr){11-19}
& \multicolumn{3}{c}{\textbf{FLIP}} & \multicolumn{3}{c}{\textbf{ALT.}} & \multicolumn{3}{c}{\textbf{CORR.}} & \multicolumn{3}{c}{\textbf{FLIP}} & \multicolumn{3}{c}{\textbf{ALT.}} & \multicolumn{3}{c}{\textbf{CORR.}} \\
\cmidrule(lr){2-4} \cmidrule(lr){5-7} \cmidrule(lr){8-10} \cmidrule(lr){11-13} \cmidrule(lr){14-16} \cmidrule(lr){17-19}
\textbf{Level} 
& Rate & $p$ & $p^*$ 
& Rate & $p$ & $p^*$ 
& Rate & $p$ & $p^*$ 
& Rate & $p$ & $p^*$ 
& Rate & $p$ & $p^*$ 
& Rate & $p$ & $p^*$ \\
\midrule
\texttt{M4}
& 0.567 & 0     & 0     
& 0.395 & 0.019 & 0.032 
& 0.608 & 0     & 0     
& 0.307 & 0     & 0     
& 0.184 & 0     & 0     
& 0.344 & 0     & 0     \\
\texttt{M8}
& 0.367 & 0.084 & 0.125 
& 0.400 & 0.013 & 0.022 
& 0.307 & 0.572 & 0.665 
& 0.127 & 0     & 0     
& 0.067 & 0.082 & 0.233 
& 0.073 & 0.042 & 0.076 \\
\texttt{M16}
& 0.333 & 0.303 & 0.389 
& 0.280 & 0.814 & 0.874 
& 0.347 & 0.194 & 0.264 
& 0.080 & 0.020 & 0.076 
& 0.053 & 0.245 & 0.513 
& 0.060 & 0.147 & 0.361 \\
\texttt{M32} 
& 0.340 & 0.245 & 0.324 
& 0.327 & 0.366 & 0.458 
& 0.340 & 0.245 & 0.324 
& 0.053 & 0.245 & 0.513 
& 0.087 & 0.009 & 0.039 
& 0.053 & 0.245 & 0.513 \\
\texttt{M64} 
& 0.373 & 0.061 & 0.093 
& 0.380 & 0.043 & 0.067 
& 0.353 & 0.150 & 0.210 
& 0.047 & 0.375 & 0.670 
& 0.053 & 0.245 & 0.361 
& 0.053 & 0.245 & 0.513 \\
\bottomrule
\end{tabular*}

\vspace{0.5em}

\caption{$\chi^2$ goodness-of-fit test results on full PLRV histogram (bin size 1, $n=150$) for controlled entropy manipulations. For each manipulation level, we report the reduced chi-squared statistic and $p$-value. $\chi^{2*}_{\text{red}}$ and $p^*$ denote values obtained after excluding the two bins at the PLRV discontinuity. Under the null hypothesis, $\chi^2_{\text{red}} \approx 1$.}
\label{tab:chi2_manipulated}
\small
\begin{tabular*}{\textwidth}{@{\extracolsep{\fill}}l*{2}{cc}*{2}{cc}*{2}{cc}}
\toprule
& \multicolumn{4}{c}{\textbf{FLIP}} & \multicolumn{4}{c}{\textbf{ALTERNATE}} & \multicolumn{4}{c}{\textbf{CORRELATION}} \\
\cmidrule(lr){2-5} \cmidrule(lr){6-9} \cmidrule(lr){10-13}
\textbf{Level} 
& $\chi^2_{\text{red}}$ & $p$ & $\chi^{2*}_{\text{red}}$ & $p^*$ 
& $\chi^2_{\text{red}}$ & $p$ & $\chi^{2*}_{\text{red}}$ & $p^*$ 
& $\chi^2_{\text{red}}$ & $p$ & $\chi^{2*}_{\text{red}}$ & $p^*$ \\
\midrule
\texttt{M4}  & 23.03 & 0     & 16.16 & 0     & 10.61 & 0     & 8.31  & 0     & 16.83 & 0     & 11.93 & 0     \\
\texttt{M8}  & 3.84  & 0     & 2.34  & 0     & 2.90  & 0     & 1.37  & 0.002 & 1.55  & 0     & 1.52  & 0     \\
\texttt{M16} & 1.94  & 0     & 1.27  & 0.015 & 1.43  & 0     & 0.89  & 0.823 & 1.06  & 0.291 & 1.00  & 0.477 \\
\texttt{M32} & 1.33  & 0.004 & 1.01  & 0.442 & 1.38  & 0.001 & 1.17  & 0.078 & 1.23  & 0.031 & 1.06  & 0.279 \\
\texttt{M64} & 1.29  & 0.009 & 1.06  & 0.293 & 1.42  & 0.001 & 1.19  & 0.054 & 1.16  & 0.093 & 1.02  & 0.409 \\
\bottomrule
\end{tabular*}

\vspace{-1em}
\end{table*}

We performed this validation under two conditions. First, with frozen entropy. We fed identical entropy sequences to both $\mathcal{D}_1$ and $\mathcal{D}_2$ queries, isolating uncertainty purely from finite sampling. The histograms appeared nearly identical, differing only by a one-bin shift, with PLRV bounded within $e^{\pm\varepsilon}$ in the central region as expected for pure differential privacy with $\varepsilon = 0.1$. The ratio $R$ clustered around 1 across bins (Figure \ref{fig:statistical_significance}), confirming our binomial uncertainty model. Counting bins where ratios deviate from 1 by more than $2\sigma$, we found exceedance rates consistent with the expected 5\% ($\texttt{REF}$: 4.97\%, $p = 0.31$; $\texttt{CSPRNG}$: 3.85\%, $p = 0.59$).

Second, we repeated validation with unfrozen conditions, allowing each of 100 repetitions to sample independently from the entropy source. Results remained consistent with the binomial model ($\texttt{REF}$: 6.63\%, $p = 0.07$; $\texttt{CSPRNG}$: 4.93\%, $p = 0.32$), demonstrating that high-quality entropy sources (both quantum TRNG and cryptographic PRNG) produce statistically indistinguishable results. This establishes a validated baseline for comparing against degraded entropy sources.

\section{Results - Entropy Quality Impact Analysis}
\label{sec:results}

Having validated our measurement framework, we evaluated how different entropy sources affect empirical privacy loss. Our analysis examines baseline sources (\texttt{REF}, \texttt{CSPRNG}, \texttt{MIS}) and controlled manipulations (\texttt{M4}--\texttt{M64}) across three manipulation types (\texttt{FLIP}, \texttt{ALTERNATE}, \texttt{CORRELATION}). We applied three complementary statistical tests to detect deviations from the theoretical PLRV distribution. Our findings proved robust across multiple configurations, indicating that detection thresholds reflect genuine properties of how entropy degradation affects DP mechanisms rather than artifacts of analysis choices.

Because the theoretical PLRV transitions continuously from $e^{\varepsilon}$ to $e^{-\varepsilon}$ between the two distribution centers, the discrete model used in the statistical tests cannot entirely capture the behavior between the two bins immediately adjacent to the transition. The empirical ratio in these bins systematically deviates from the model prediction, producing a baseline $\chi^2$ inflation that affects all entropy sources. We therefore report results both with and without these two bins: values marked with an asterisk ($p^*$, $\chi^{2*}_{\text{red}}$) exclude them, isolating deviations due to entropy degradation from this geometric artifact. As shown below, this distinction primarily affects baseline calibration and borderline cases. For strong manipulations, detection remains strong regardless. 

Additionally, for the strongest manipulations (\texttt{M4}), some tail bins received zero counts in one or both histograms and were excluded from the analysis, reducing the effective number of bins below 150 (see Table \ref{tab:sign_test_manipulation}).

\begin{figure*}[htp]
	\centering
	\begin{subfigure}[b]{0.45\textwidth}
		\centering
		\includegraphics[width=\textwidth]{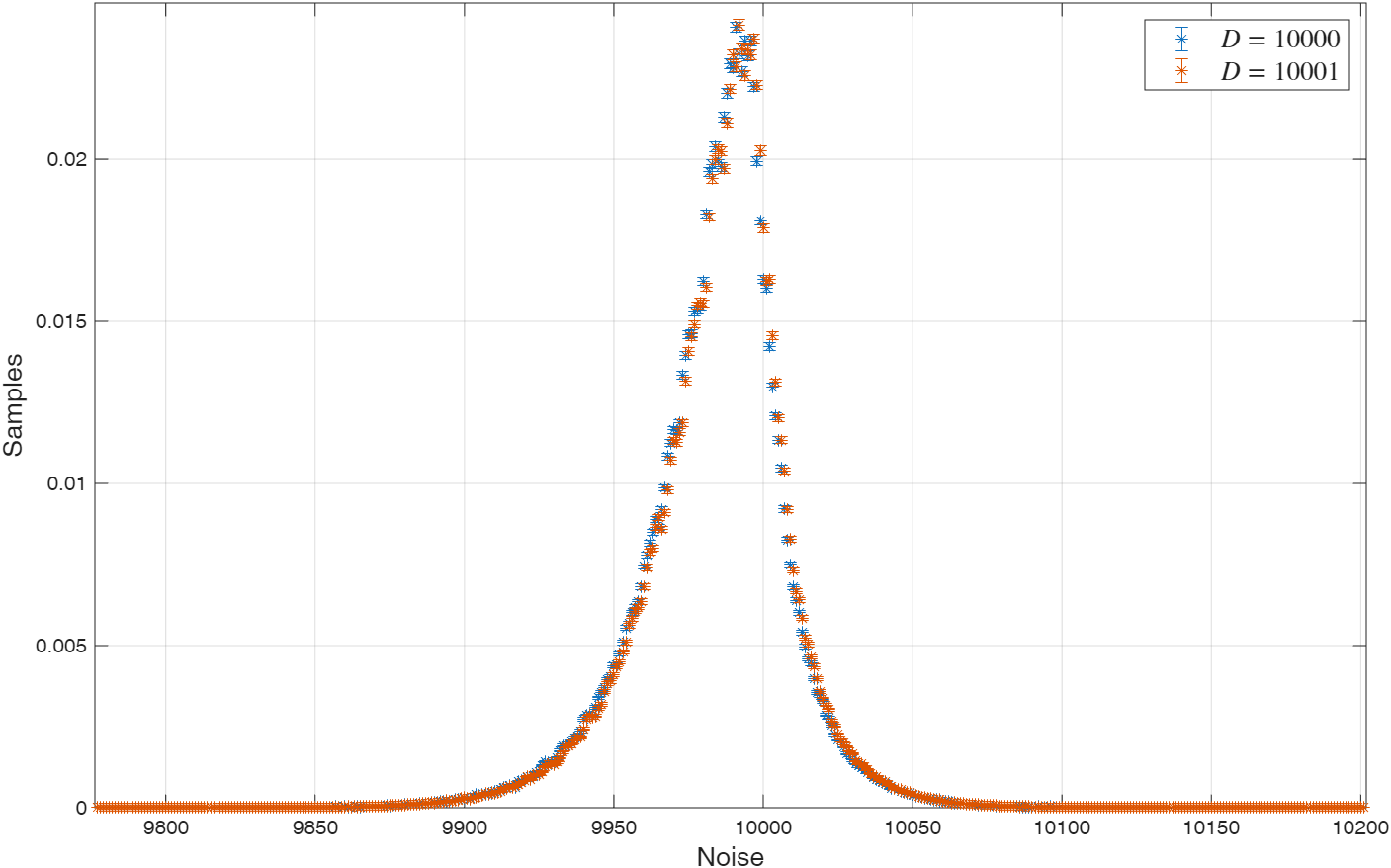}
		\caption{Distribution for \texttt{M4 SET} manipulation.}
		\label{fig:m4_set_dist}
	\end{subfigure}
	\hfill
	\begin{subfigure}[b]{0.45\textwidth}
		\centering
		\includegraphics[width=\textwidth]{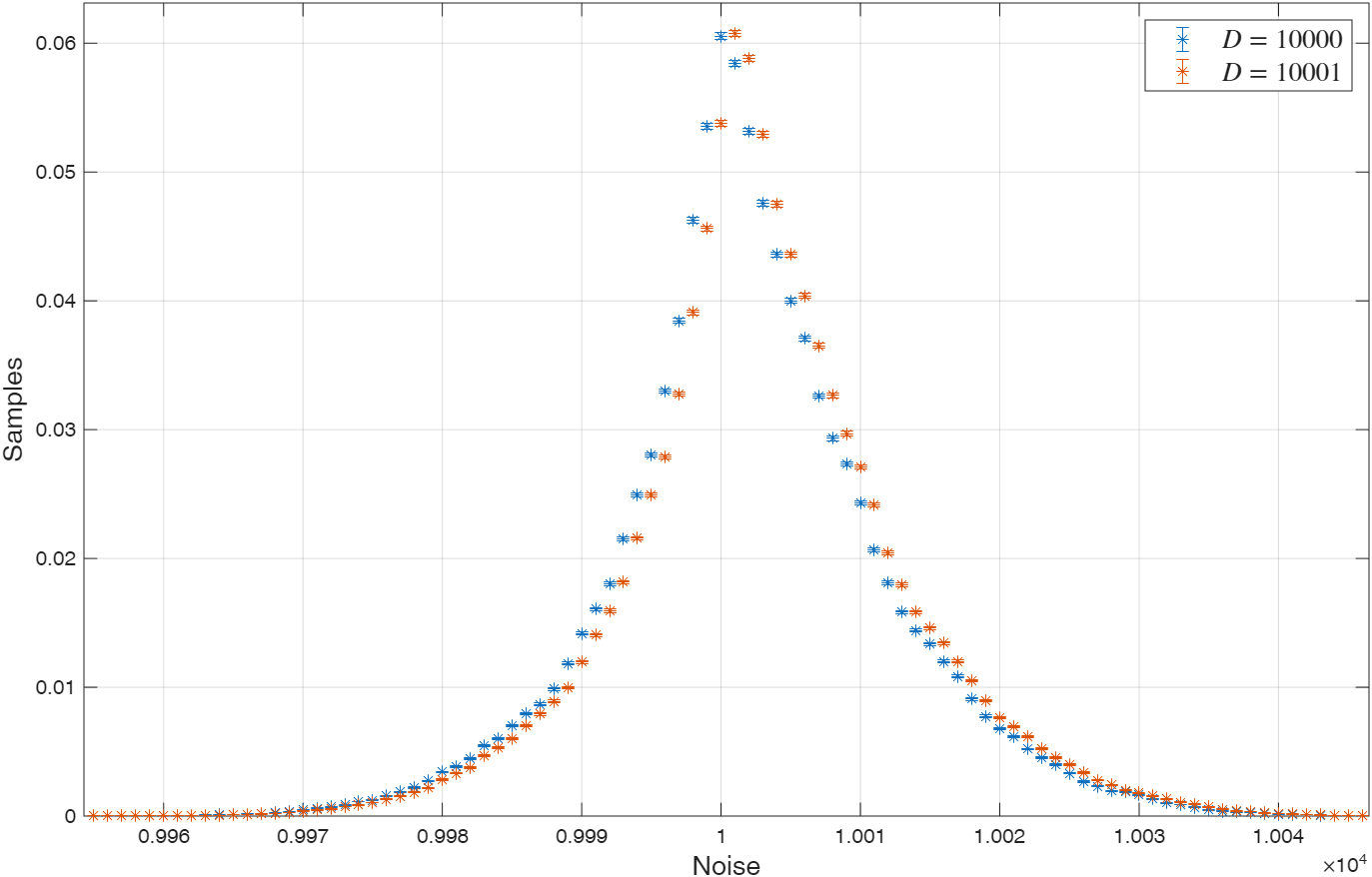}
		\caption{Distribution for \texttt{M4 CLEAR} manipulation.}
		\label{fig:m4_clear_dist}
	\end{subfigure}
	\begin{subfigure}[b]{0.45\textwidth}
		\centering
		\includegraphics[width=\textwidth]{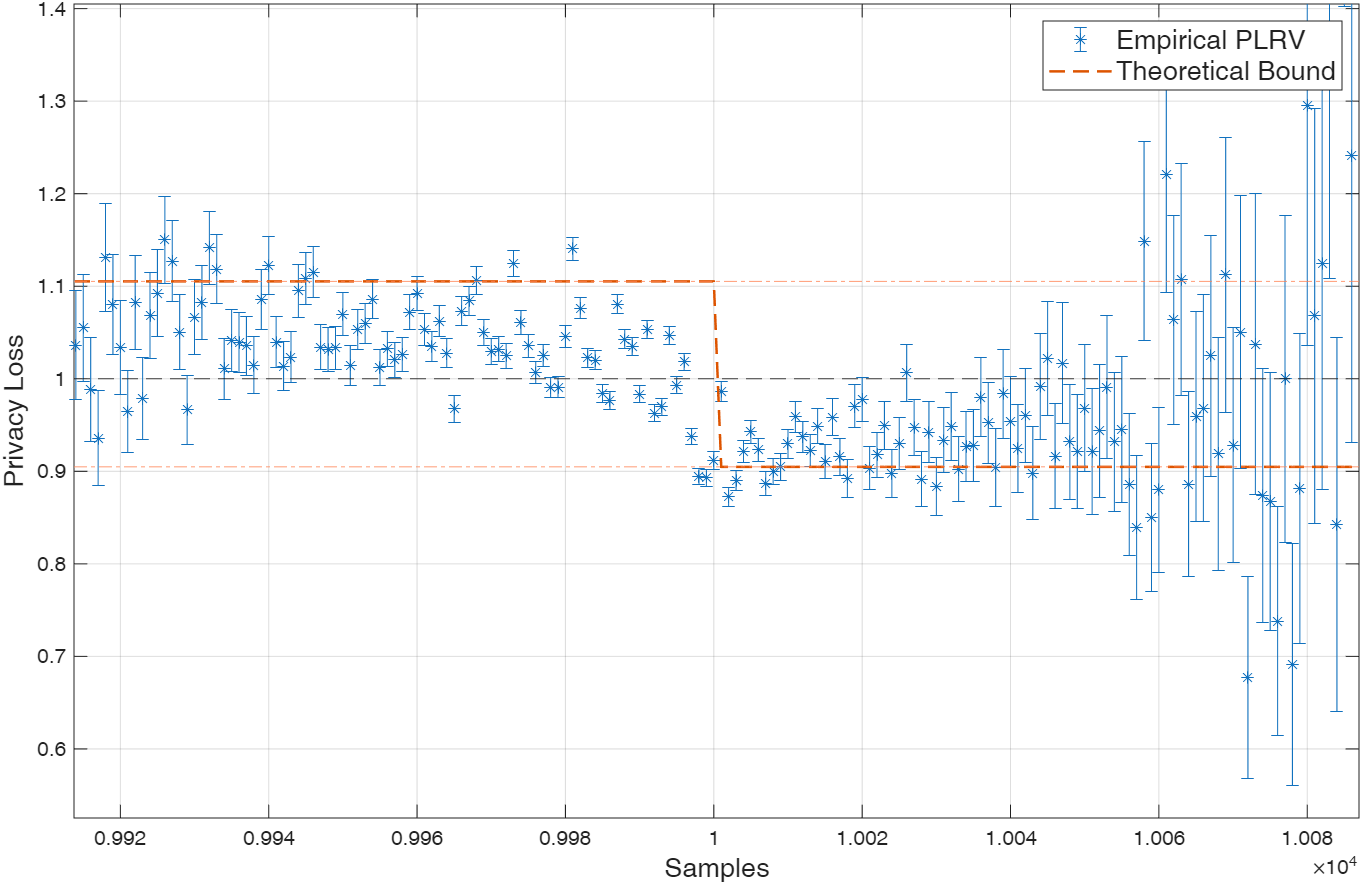}
		\caption{PLRV values for \texttt{M4 SET} manipulation.}
		\label{fig:m4_set_plrv}
	\end{subfigure}
	\hfill
	\begin{subfigure}[b]{0.45\textwidth}
		\centering
		\includegraphics[width=\textwidth]{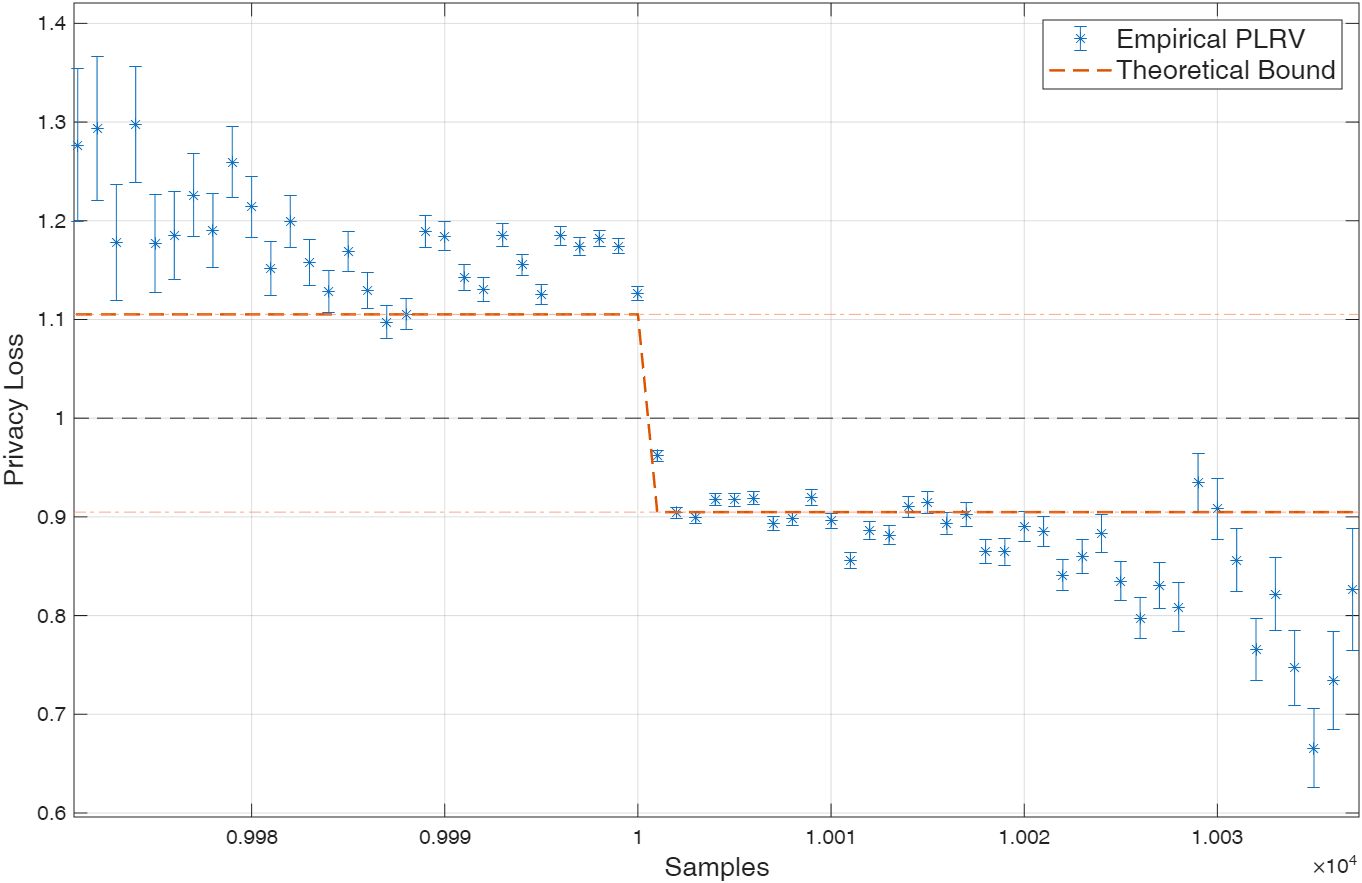}
		\caption{PLRV values for \texttt{M4 CLEAR} manipulation.}
		\label{fig:m4_clear_plrv}
	\end{subfigure}
    \caption{Output distributions and PLRV values for \texttt{M4 SET} and \texttt{CLEAR} manipulations ($\varepsilon = 0.1$, $N = 10^6$, bin size $=1$). Both produce non-Laplacian distributions (a,b), but only \texttt{CLEAR} violates PLRV bounds (d), while \texttt{SET} maintains values within constraints (c), demonstrating that statistical deviations do not necessarily imply privacy violations.}
	\label{fig:clear_set_m4_comparison}

    \vspace{-1.5em}
\end{figure*}

\subsubsection{Sign Test}

The sign test detects systematic directional biases by examining whether deviations from the theoretical model, namely the $\pm\varepsilon$ bounds, are symmetric around zero. Under the null hypothesis that the mechanism provides $\varepsilon$-DP with high-quality randomness, positive and negative deviations should occur with equal probability, following a Binomial$(n, 0.5)$ distribution.

For baseline sources (\texttt{REF}, \texttt{CSPRNG}, \texttt{MIS}), all p-values exceeded standard significance thresholds, indicating no detectable asymmetry in privacy loss deviations (Table \ref{tab:sign_test_baseline}). 

Strong manipulations (\texttt{M4}, \texttt{M8}) produced extremely small p-values across most manipulation types (Table \ref{tab:sign_test_manipulation}), signaling clear directional bias. The \texttt{FLIP}, and \texttt{CORRELATION} manipulations were particularly detectable, with the sign test rejecting the null hypothesis.

At moderate manipulation levels (\texttt{M16}), detection became less reliable, with some manipulation types showing both significant and non-significant p-values. Mild manipulations (\texttt{M32}, \texttt{M64}) mostly remained undetected, with p-values returning to non-significant ranges, establishing a clear sensitivity gradient: the sign test reliably detects strong, frequent manipulations but loses power for sparse perturbations.

Notably, \texttt{ALTERNATE} manipulations behaved differently, producing non-significant p-values even at \texttt{M4} rather than strong rejections, suggesting that alternating pattern manipulations partially cancel in their effect on directional bias.

\subsubsection{Sigma Exceedance Test}
The sigma exceedance test examines both the frequency and magnitude of deviations by computing z-scores: $z_i = (\text{estimate}_i - \text{theory}_i)/\sigma_i$. Under the $\varepsilon$-DP model, these should behave like standard normal fluctuations, with approximately 31.7\% of bins exceeding $1\sigma$ and 4.6\% exceeding $2\sigma$ purely by chance. 

For baseline sources, the analysis showed adherence to the model with exceedance rates close to expected values and non-significant p-values (Table \ref{tab:ksigma_baseline}).

For strong manipulations (\texttt{M4}), both $1\sigma$ and $2\sigma$ exceedance rates increased (Table \ref{tab:ksigma_manipulated}), with over 55\% of bins exceeding $1\sigma$ and up to 30\% exceeding $2\sigma$, occurring consistently across manipulation types, entropy sources, and with or without jump-bin exclusion. At moderate levels (\texttt{M8}), $2\sigma$ exceedances remained well above standard rates, with low p-values for \texttt{FLIP}, borderline detection for \texttt{ALTERNATE} and \texttt{CORRELATION}.

At \texttt{M16}, detection became ambiguous. \texttt{FLIP} continued producing significant excess $2\sigma$ outliers, while \texttt{ALTERNATE} showed borderline significance that did not survive jump-bin exclusion, and \texttt{CORRELATION} remained within expected ranges. Mild manipulations (\texttt{M32}, \texttt{M64}) fell below test sensitivity, with exceedance rates fluctuating near nominal values and non-significant p-values across all configurations.

\subsubsection{Chi-Squared Goodness-of-Fit Test}
The $\chi^2$ test quantifies overall fit to the theoretical PLRV via the reduced chi-squared statistic: $\chi^2_{\text{red}} = \frac{1}{n_{\text{dof}}} \sum_{i} (\text{observed}_i - \text{expected}_i)^2/\sigma_i^2$. Under the null hypothesis, $\chi^2_{\text{red}} \approx 1$.

Baseline sources showed $\chi^2_{\text{red}}$ slightly above 1 (Table \ref{tab:chi2_baseline}), with \texttt{REF} at 1.30 ($p = 0.008$) and \texttt{CSPRNG} at 1.31 ($p = 0.007$), formally rejecting the null hypothesis. However, excluding the two transition bins brought both baselines to expected values: $\chi^{2*}_{\text{red}} = 1.11$ ($p^* = 0.174$) for \texttt{REF} and $1.09$ ($p^* = 0.207$) for \texttt{CSPRNG}. The miscalibrated QRNG (\texttt{MIS}), which was already non-significant at baseline ($\chi^2_{\text{red}} = 1.15$, $p = 0.102$), dropped further to $\chi^{2*}_{\text{red}} = 0.95$ ($p^* = 0.667$). This confirms that the original inflation was caused by the discrete transition at the PLRV midpoint rather than by model failure.

Strong manipulations (\texttt{M4}, \texttt{M8}) caused strong rejections with $\chi^2_{\text{red}}$ values between 3 and 23 and zero p-values across all entropy sources (Table \ref{tab:chi2_manipulated}). Jump-bin exclusion reduced these values (e.g., from \texttt{FLIP M4} $23.0$ to $16.2^*$) but detection remained strong, confirming that the distortion is distributed across the full histogram.

At \texttt{M16}, the jump-bin exclusion changed the picture substantially. \texttt{FLIP} manipulations remained detectable ($\chi^{2*}_{\text{red}} \approx 1.27$, $p^* \approx 0.015$), but \texttt{ALTERNATE} went from rejected ($\chi^2_{\text{red}} = 1.43$, $p < 10^{-3}$) to fully baseline-consistent ($\chi^{2*}_{\text{red}} = 0.89$, $p^* = 0.823$), indicating that the rejection was due to the transition bins rather than entropy degradation. \texttt{CORRELATION} was baseline-consistent regardless ($p > 0.29$, $p^* > 0.48$).

At \texttt{M32}, a similar pattern emerged: \texttt{FLIP} went from formal rejection ($p = 0.004$) to baseline-consistent ($p^* = 0.442$), establishing that these mild manipulations fall below the test's sensitivity. At \texttt{M64}, all manipulation types were indistinguishable from baseline across all entropy sources.

\subsubsection{Statistical Detection vs. Privacy Violation}

Statistical deviations do not necessarily imply privacy guarantee violations. Figure \ref{fig:clear_set_m4_comparison} illustrates this distinction through \texttt{M4 CLEAR} and \texttt{M4 SET} manipulations, two complementary manipulations that deterministically modify 1 bit every 4 bits. Both strongly fail all statistical tests and yield non-Laplacian output distributions (Figures \ref{fig:m4_set_dist} and \ref{fig:m4_clear_dist}). However, only \texttt{CLEAR} produces PLRV values that violate expected privacy bounds (Figure \ref{fig:m4_clear_plrv}), indicating privacy loss. The \texttt{SET} manipulation, despite being statistically detected as non-Laplacian, yields PLRV values consistently within privacy constraints (Figure \ref{fig:m4_set_plrv}). This demonstrates that detecting distributional anomalies requires further scrutiny to determine whether empirical PLRV actually exceeds theoretical thresholds. These results establish that while statistical tests can detect entropy degradation, the relationship between statistical detection and actual privacy violations is nuanced and depends on how the degradation manifests in the PLRV distribution.

\section{Conclusions}
\label{sec:conclusions}

Our statistical framework reliably detects deviations when approximately 1 bit in every 8 to 16 is manipulated. Beyond this threshold, detection becomes increasingly ambiguous, particularly for correlation-based perturbations. The three tests provide complementary information: the sign test detects directional biases but misses amplitude changes; the sigma exceedance test captures both frequency and magnitude of exceedances; the $\chi^2$ test assesses overall distributional shape.

An important asymmetry emerged in how entropy degradation types affect DP mechanisms. Differential privacy proves remarkably resilient to correlations from miscalibrations (\texttt{MIS}) and low-frequency correlations (\texttt{CORRELATION}), which often remain undetected or produce PLRV values within acceptable bounds even when output distributions deviate from ideal Laplace shape. In contrast, bit-level biases (\texttt{FLIP}) distort the sampling process more directly, producing distributions that fail statistical tests and can violate privacy guarantees.

Excluding the two bins at the PLRV transition resolved baseline $\chi^2$ inflation across all entropy sources, confirming that this inflation was a geometric artifact due to the discretization of the model rather than a sign of model failure. Bit-level biases more strongly distort bit-streams and are easier to detect, while temporal correlations affect sequence structure more subtly. These results suggest that while DP mechanisms are robust to imperfect randomness, users requiring strict adherence to mathematical bounds should ensure high-quality entropy sources.

\section*{Acknowledgements}
This project was partially funded by the European Union’s ``Next Generation EU" Program under the Piano Nazionale di Ripresa e Resilienza (PNRR) through NQSTI Spoke8.

\bibliographystyle{unsrt}
\bibliography{bibliography}

\end{document}